\documentclass[aps,prb,superscriptaddress,twocolumn,no-footinbib,showpacs,floatfix,amsmath,amssymb]{revtex4-1}
 
\usepackage{graphicx}
\usepackage{dcolumn}
\usepackage{bm}
\usepackage[normalem]{ulem}
\usepackage{subfigure} 
\usepackage{soul,xcolor}
\usepackage{siunitx}
\usepackage{textcomp}
\setstcolor{red}
\RequirePackage{color}

\usepackage[capitalise,nameinlink,noabbrev]{cleveref}
\crefname{equation}{}{}
\crefname{enumi}{}{}

\begin{document}


\title{Valley engineering by strain in Kekul\'{e}-distorted graphene }
\author{Elias Andrade}
\affiliation{%
 Facultad de Ciencias, Universidad Aut\'{o}noma de Baja California,
Apdo. Postal 1880, 22800 Ensenada, Baja California, M\'{e}xico. 
}%
\author{Ramon Carrillo-Bastos}
\email{ramoncarrillo@uabc.edu.mx}
\affiliation{%
 Facultad de Ciencias, Universidad Aut\'{o}noma de Baja California,
Apdo. Postal 1880, 22800 Ensenada, Baja California, M\'{e}xico. 
}%
\author{Gerardo G. Naumis}
\email{naumis@fisica.unam.mx}
 \homepage{\\http://www.fisica.unam.mx/personales/naumis/}
\affiliation{
 Depto. de Sistemas Complejos, Instituto de F\'{i}sica, Universidad Nacional Aut\'{o}noma de M\'{e}xico (UNAM). Apdo. Postal 20-364, 01000 M\'{e}xico D.F., M\'{e}xico
 }%

\date{\today}

\begin{abstract}
A Kekul\'e bond texture in graphene modifies the electronic band structure by folding the Brillouin zone and bringing the two inequivalent Dirac points to the center. This can result, in the opening of a gap (Kek-O) or the locking of the valley degree of freedom with the direction of motion (Kek-Y). We analyze the effects of uniaxial strain on the band structure of Kekul\'e-distorted graphene for both textures. Using a tight-binding approach, we introduce strain by considering the hopping renormalization and corresponding geometrical modifications of the Brillouin zone. We numerically evaluate the dispersion relation and present analytical expressions for the low-energy limit. Our results indicate the emergence of a Zeeman-like term due to the coupling of the pseudospin with the pseudomagnetic strain potential which separates the valleys by moving them in opposite directions away from the center of the Brillouin zone. For the Kek-O phase, this results in a competition between the Kekul\'e parameter that opens a gap and the magnitude of strain which closes it. While for the Kek-Y phase,  in a superposition of two shifted Dirac cones. As the Dirac cones are much closer in the supperlattice reciprocal space that in pristine graphene, we propose strain as a control parameter for intervalley scattering. 
\end{abstract}

\maketitle

\section{\label{sec:intro}Introduction}

In graphene, the electronic properties are dominated by the two inequivalent local minima in the conduction band, located at the high symmetry Brillouin zone points  $\bm{K_{+}}$ and $\bm{K_{-}}$, and referred to the $K_D^+$ and $K_D^-$ valley, respectively. This endows  low-energy electrons with an additional degree of freedom\cite{Schaibley}, known as valley isospin. In pristine membranes\cite{neto2009electronic}, these two valleys have gapless Dirac spectra, which are degenerate in energy, related by time-reversal symmetry, and well separated in reciprocal space by the Kekul\'e vector $\bm{G}=\bm{K_{+}}-\bm{K_{-}}$. However, if graphene is subject to a periodic perturbation, with a spatial periodicity associated with $\bm{G}$ (Kekul\'{e} distorsion), a superlattice with a tripled unit cell (of the size of a hexagonal ring) is formed. As a consequence, the two Dirac cones at opposite corners ($K_D^+$ and $K_D^-$) are folded onto the center $\Gamma$ of the new hexagonal superlattice Brillouin zone\cite{Chamon2007}. Almost twenty years ago, Claudio Chamon showed that a bond distortion  mimicking the Kekul\'e structure for benzene (Kek-O) provides such a periodicity in graphene, which opens a gap by mixing the two valley species\cite{Chamon2000}. Interestingly, graphene with a Kek-O distortion is also expected to show topological charge fractionalization\cite{Chamon2007}, and other topological properties\cite{wakabayashi,wu2016}. 

Although experimentally achievable in analogues of graphene\cite{manoharan,li-phonons}, up to now the Kek-O phase in graphene has not become a physical reality. Nevertheless, theoretical studies suggest that the Kek-O phase can be obtained by depositing graphene on top of a topological insulator\cite{ontop}, by applying uniaxial strain\cite{Sorella} or by placing atoms adsorbed on its surface\cite{cheianov}.
The latter proposal was pursued by Gutierrez \textit{et al.}\cite{Gutierrez}, who experimentally found another Kekul\'{e} distorsion, the Kekul\'e-Y (Kek-Y) phase, which consists of a periodic modification of the three bonds (in form of the letter Y) surrounding one of the atoms of the new hexagonal unit cell. Recently, Gamayun \textit{et al} showed that this Kek-Y bond texture results in the locking of valley isospin with the direction of motion (momentum), breaking the valley degeneracy while preserving the massless character of the Dirac fermions. This effect opens a new way to control the valley degree of freedom in graphene\cite{Gamayun}.  

There have been several theoretical proposals to manipulate the valley degree of freedom in graphene\cite{rycerz2007valley,Fujita2010,guinea2013,wang2014,grujic2014,ren2015,carrillo2016strained,stegmann2016valley,jones2017quantized,wu2016full,Cazalilla,Asmar-minimal,Luo2017,Beenakker,Yee2017,Brown2018,settnes2017valley,milovanovic2016strained,Roche,carrillo2018enhanced,stegmann2018,zhai2018local,ValleyPRL2018}, including the celebrated Valley Hall effect\cite{Xiao2007} produced by Berry curvature\cite{xiao-review}, and the  use of strain\cite{AMORIM20161,NaumisReview,vozmediano2010gauge}. The former has been recently  observed in graphene superlattices by nonlocal transport measurement\cite{gorbachev2014,komatsu2018}. The effects of the latter\cite{Crommie,Theory-LL01,Theory-LL02} are strong, measurable and expected to be valley asymmetric\cite{settnes2017valley,milovanovic2016strained,Roche,carrillo2018enhanced}. In fact, both couple asymmetrically with each valley by breaking the inversion symmetry while preserving time reversal. Nevertheless, strain offers the advantage of being tunable and it is in intimate relation with the kekul\'{e} phase, since this phase is  expected to appear in the presence of uniaxial strain\cite{Sorella,guinea}. In general, uniaxial strain alters the band structure of graphene by (1) distorting the shape of the Brillouin zone, thus changing the geometrical position of the high symmetry points due to the modification of the lattice vectors\cite{oliva-Leyva-PRB}, and (2) moving the Dirac cones away from the high symmetry points, since it changes unevenly the three hopping energies connecting neighboring sites\cite{Pereira2009}. These two effects should be taken into account to obtain the low energy approximation for graphene, otherwise unphysical results are obtained even in the simplest cases\cite{oliva-Leyva-PRB,oliva-leyva-physA}. 

Inspired by the results described above, in the present manuscript we evaluate the effect of uniaxial strain on Kekul\'{e} distorted graphene in both phases: Kek-O and Kek-Y. Using the tight binding approximation, we write the Hamiltonian for Kekul\'e-distorted graphene and introduce strain by changing the hopping integrals and atomic positions in the lattice. The layout of this paper is the following: In Section II, we present the model, as well as the resulting band structures. Section III is devoted to obtaining a low-energy effective Hamiltonian; and in Section IV, we provide the final conclusions and remarks.

\section{Hamiltonian for strained Kekul\'{e} distorted graphene}\label{Sec_II}

\begin{figure}[!htbp]
		\begin{center}
		\includegraphics[width=8.5cm]{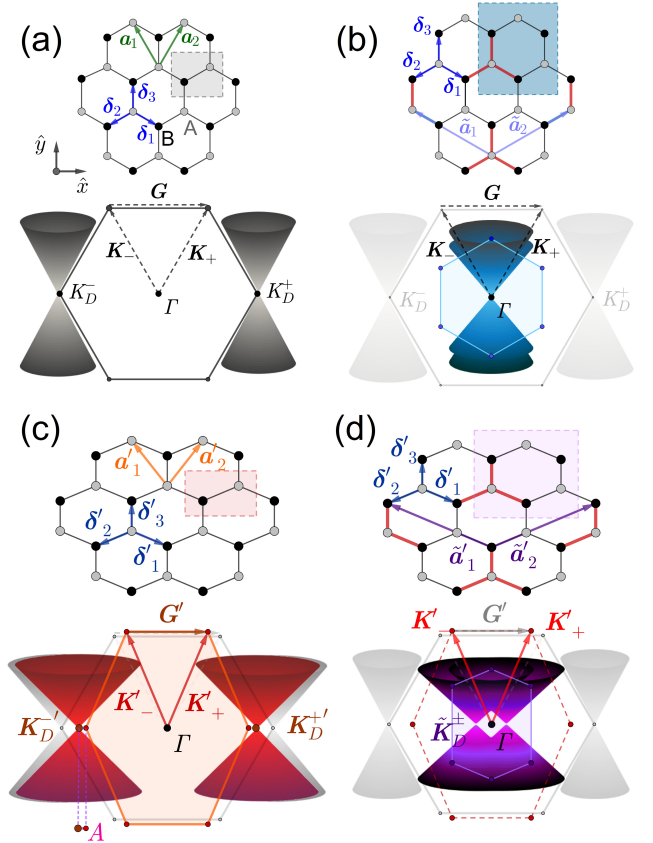}
		\end{center}         
      \caption{Lattices and Brillouin Zones for: (a) pristine graphene, (b) Kek-Y distorted graphene (red bonds),  (c) strained graphene, and (d) Kek-Y distorted graphene with strain. The Kekul\'e and strained Kekul\'e vectors $\bm{G}$ and  $\bm{G}'$ are indicated in the upper side of the hexagonal Brillouin zones for each case.  In a), the Dirac cones for pristine graphene are indicated in dark gray. In b), the gray Dirac cones are folded into a degenerate Dirac cone (in blue) to the $\Gamma$ point, while in c), the original Dirac cones (in gray) are deformed and translated to the points $\bm{K}_D^{\pm}$, as indicated in red.  In d), the Dirac cones (gray) are folded into the $\Gamma$ point, but the strain breaks the degeneracy resulting in the overlapping of two shifted Dirac cones, both indicated in purple.}
		\label{fig:NuevaBrillouin}
\end{figure}

Let us start by considering  a pure Kekul\'e pattern on unstrained graphene.  The electronic properties are well described  by a tight-binding Hamiltonian for a single $\pi$-orbital per carbon site \cite{Gamayun}, 
\begin{equation}
H=- \sum_{\bm{r}} \sum_{j=1}^{3} t_{\bm{r},j} a_{\bm r}^{\dagger}b_{\bm r+ \bm \delta_j}+H.c.,
\end{equation}
where $\bm{r}$ runs over the atomic positions of graphene's sublattice A, given by $\bm{r} = n_1\bm{a}_1 + n_2\bm{a}_2$, with $n_1$ and $n_2$ integers. The lattice vectors are $\boldsymbol{a}_1=a(-\frac{\sqrt{3}}{2},\frac{3}{2})$, and $\boldsymbol{a}_2=a(\frac{\sqrt{3}}{2},\frac{3}{2}) $, with $a= 1.42$~\AA. Each vector $\bm{\delta}_i$  points to one of the three nearest-neighbor sites belonging to sublattice B, and surrounding the site located at a given $\bm{r}$, as shown in Fig. \ref{fig:NuevaBrillouin} $[\bm{\delta}_1=a(\frac{\sqrt[]{3}}{2},-\frac{1}{2})$, $\bm{\delta}_2=-a(\frac{\sqrt[]{3}}{2},\frac{1}{2})$, and $\bm{\delta}_3=a(0,1)]$. The set of tight-binding parameters describing the  bond-density wave of the Kekul\'e pattern is given by\cite{Gamayun}
\begin{equation}\label{tKekule}
\begin{split}
t_{\bm{r},j}=t_0 \Big[&1+\Delta e^{i(p \bm K_+ + q \bm K_-) \cdot \bm \delta_j + i \bm G \cdot \bm r}  \\
&+ \Delta^* e^{-i(p \bm K_++q \bm K_-) \cdot \bm \delta_j - i \bm G \cdot \bm r}\Big],
\end{split}
\end{equation}  
where $t_0\approx 2.7 eV$ is the hopping-parameter for pristine graphene, $\Delta=e^{i2\pi m/3}\Delta_0$ is the Kekul\'{e} coupling with amplitude $\Delta_0$ and $m$ an arbitrary integer number, $\bm{K}_{\pm}=\frac{2 \pi}{9a}\sqrt{3}(\pm 1,\sqrt{3})$ are the high-symmetry points of graphene such that the  Kekul\'e wave vector is $\bm G=\frac{4 \pi}{9a}\sqrt{3}(1,0)$. Given that $p$ and $q$ are integers, the value of the Kekul\'e-distorted hopping-parameter $t_{\bm{r},j}$ oscillates in space  
between the values $t_0(1-\Delta_0)$  and $t_0(1+2\Delta_0)$, generating  a Kekul\'e texture accordingly to the index \cite{Gamayun}
\begin{equation}
\nu = 1+q-p \mod 3,
\label{Eq:Nu}
\end{equation}
with a Kek-O texture for $\nu=0$, and Kek-Y for $\nu=\pm1$.

Let us now consider the effects of strain. When a strain field $\boldsymbol{u}=(u_{x}(x,y),u_{y}(x,y))$ is applied to pristine graphene, 
the atomic positions $\bm{r}$ change to,
\begin{equation}\label{transform}
\boldsymbol{r}'=(1 + \bar{\epsilon}) \cdot \boldsymbol{r},
\end{equation}
where $\bar{\epsilon}$ is the strain tensor  with components \cite{NaumisReview},
\begin{equation}
\epsilon_{ij}=\frac{1}{2}\left(\frac{\partial u_j}{\partial x_i}+\frac{\partial u_i}{ \partial x_j} \right).
\end{equation}
where $i=x,y$ and $j=x,y$. The local distance between neighbor atoms gets modified accordingly \cite{NaumisReview},
\begin{equation}
\boldsymbol{\delta}_j'=(1 + \bar{\epsilon}) \cdot \boldsymbol{\delta_}j,
\end{equation}
and similarly the basis vectors.
\begin{equation}
\boldsymbol{a}_j'=(1 + \bar{\epsilon}) \cdot \boldsymbol{a}_j,
\end{equation}
as seen in Fig. \ref{fig:NuevaBrillouin}.

Notice that the considered strain is uniform and thus space independent. This case also serves as a first approximation for smooth strain profiles. As the strain is uniform, it can be written as follows, 
\begin{equation}
\bar{\epsilon}=\left[ {\begin{array}{cc}
   \epsilon_{xx} & \epsilon_{xy} \\
   \epsilon_{xy} & \epsilon_{yy} \\
  \end{array} } \right]=
  \left[ {\begin{array}{cc}
   \epsilon_Z & \epsilon_{S} \\
   \epsilon_{S} & \epsilon_A \\
  \end{array} } \right].
\end{equation}
In the previous expression, the space-independent parameters ${{\epsilon}_{A}}$ and ${{\epsilon}_{Z}}$ denote uniaxial strain applied along the zigzag and armchair directions, respectively, and $\epsilon_{S}$ is the shear strain.
This tensor can be parametrized in terms of $\epsilon$ (the magnitude of the applied strain), its angular direction $\theta$ (with respect to the $x$-axis), and $\rho$, the Poisson ratio which relates the strain components with a value of $\rho=0.165$ for graphene\cite{Botello2018},
\begin{equation}
\bar{\epsilon}=\left[ {\begin{array}{cc}
   \epsilon (\cos^2 \theta - \rho \sin^2 \theta) & \epsilon (1+\rho)\cos\theta \sin \theta \\
   \epsilon (1+\rho)\cos\theta \sin \theta & \epsilon (\sin^2 \theta - \rho \cos^2 \theta) \\
  \end{array} } \right].
\end{equation}
In the absence of a Kekul\'e pattern, the tight-binding parameter for the strained lattice is  given by \cite{NaumisReview}
\begin{equation}\label{tdej}
t_{j}=t_0 e^{-\beta \left(\frac{|\bm{\delta}_j'|}{a}-1\right)},
\end{equation}
where $\beta$ is the Gruneissen parameter, estimated to be $\beta \approx 3$ for graphene. It is important to remark that second- and third-neighbor interactions are always present, which depend upon the bond torsion-angle \cite{Botello2018}.  These effects will be  neglected here as a first approximation.

Now, we can combine strain with a Kekul\'e pattern as follows: First, due to the modified distance between sites, the tight-binding parameter $t_0$ in Eq.~\cref{tKekule}  is replaced by $t_j$, defined by Eq.~\cref{tdej}. Second, we need to keep the Kekul\'e density-wave bond ordering. Thus, we can proceed by observing that the phases of the pattern are preserved if\cite{Gamayun},  
\begin{equation}\label{phaseinvariance}
(p\bm{K}_{+}+q\bm{K}_{-}) \cdot \bm{\delta}_j+  \bm{G} \cdot \bm{r}=(p\bm{K}_{+}'+q\bm{K}_-') \cdot \delta_j' +  \bm{G}' \cdot \bm{r}',
\end{equation}
as long as we define,
\begin{subequations}
\begin{equation}
 \bm K_\pm'= (1+\bar{\epsilon})^{-1} \cdot \bm K_\pm,
\end{equation}
\begin{equation}
 \bm G'= \bm K_+' - \bm K_-' = (1+  \bar{\epsilon})^{-1} \cdot  \bm G.
\end{equation}
\end{subequations}
this constitutes a systematical procedure to introduce uniaxial strain to graphene supperlattices. 

Although it is tempting to think  of $\bm{K}_{\pm}'$ as the reciprocal transformation of Eq. \cref{transform} for the high-symmetry points of the deformed lattice,  in general, it turns out that the transformed reciprocal vectors of the high symmetry points of pristine graphene do not coincide with the high-symmetry points of the deformed lattice Brillouin zone\cite{Pereira2009, NaumisReview}. Moreover, strain changes  the symmetry of the Bravais lattice. The high-symmetry points of the first Brillouin zone strained lattice must be labeled differently. As an example, the  $\bm{K}_{\pm}$ points of the P6/mmm
space group  after a uniaxial  strain  are replaced by the $\bm{F}_0$ and  $\bm{\Delta}_0$ points in  the  Cmmm space  group \cite{NaumisReview}. Also, we stress out that  Dirac points $\bm{K}_{D}^{'\pm}$ corresponding to the deformed lattice energy dispersion do not necessarily coincide neither with $\bm{K}_{\pm}'$ nor with $\bm{F}_0$ or  $\bm{\Delta}_0$. Fig. \ref{fig:NuevaBrillouin} brings a sketch of these general observations, and serves as a warning to avoid confusions about such aspects\cite{NaumisReview}. 

The modification of the tunneling parameter [Eq. \cref{tKekule}] and the change of the pattern phases [Eq. \cref{phaseinvariance}] caused by strain result in a new set of tight-binding parameters $\tilde{t}_{\bm{r}',j}$,
\begin{equation}\label{tStrainKek}
\begin{split}
\tilde{t}_{\bm{r}',j}=t_{j}\Big[&1+\Delta e^{i(p\bm{K}_+'+q\bm{K}_-') \cdot \bm{\delta}_j' + i \bm{G}' \cdot \bm{r}'} \\
&+ \Delta^* e^{-i(p\bm{K}_+'+q\bm{K}_-') \cdot \bm{\delta}_j' - i \bm{G}' \cdot \bm{r}'}\Big].
\end{split}
\end{equation}
Therefore the new Hamiltonian for the applied strain on a Kekul\'{e} pattern is the following,
\begin{equation}\label{HKek}
H=- \sum_{\bm{r}'} \sum_j \tilde{t}_{\bm{r}',j} a_{\bm{r}'}^{\dagger}b_{\bm{r}'+\bm{\delta}_j'}+H.c..
\end{equation}
Such a Hamiltonian can be written in reciprocal space by taking a Fourier transform of the anhilation/creation operators. The three terms  in Eq.~\cref{tStrainKek}  lead to Hamiltonian $H(\boldsymbol{k})=H_1(\boldsymbol{k})+H_2(\boldsymbol{k})+H_3(\boldsymbol{k})$, where $H_1(\boldsymbol{k})$ is the contribution from the Fourier transform that arises from Eq. \cref{HKek} by considering the first term in Eq. \cref{tStrainKek}: 
\begin{equation}
\begin{aligned}
H_1(\boldsymbol{k})&=-\frac{1}{2 \pi} \sum_{\bm{r}',j} \int_{\bm{k},\bm{k''}} 
t_j a_{\boldsymbol{k}''}^\dagger b_{\boldsymbol{k}} e^{i\boldsymbol{k} \cdot \boldsymbol{\delta}_j'}e^{i\boldsymbol{r}' \cdot (\boldsymbol{k}-\boldsymbol{k}'')}d^{2}k d^{2}k'',\\
&=-\int_{\boldsymbol{k}} \sum_j  t_j a_{\boldsymbol{k}}^\dagger b_{\boldsymbol{k}} e^{i\boldsymbol{k} \cdot \boldsymbol{\delta}_j'} d^{2}k,\\
&=-\int_{\boldsymbol{k}} s'(\boldsymbol{k}) a_{\boldsymbol{k}}^\dagger b_{\bm{k}} d^{2}k,
\end{aligned}
\end{equation}
where $s'(\boldsymbol{k})$ is the  dispersion relation for strained graphene  without the Kekul\'e pattern,
\begin{equation}
s'(\boldsymbol{k})=\sum_j  t_j e^{i \boldsymbol{k} \cdot \boldsymbol{\delta}_j'}.
\end{equation}
\\
The second term, $H_2(\boldsymbol{k})$, is
\begin{widetext}
\begin{equation}
\begin{aligned}
H_2(\boldsymbol{k})&=-\frac{\Delta}{2 \pi} \sum_{r',j}  \int_{\bm{k},\bm{k}''} t_j a_{\bm{k}''}^\dagger b_{\bm{k}} e^{i(\bm{K}_+'+q\bm{K}_{-}')\cdot \bm{\delta}_j'}e^{-i\bm{r}' \cdot (\bm{k}''-[\bm{k}+\bm{G}'])}d^{2}k d^{2}k'',\\
&=-\Delta \int_{\bm{k}} \sum_j  t_j a_{\bm{k}+\bm{G}'}^\dagger b_{\bm{k}} e^{i(\bm{k}+p\bm{K}_+'+q\bm{K}_-')\cdot \bm{\delta}_j'} d^{2}k,\\
&=-\Delta \int_{\bm{k}} s'(\bm{k}+p\bm{K}_{+}'+q\bm{K}_-') a_{\bm{k}+\bm{G}'}^\dagger b_{\bm{k}}  d^{2}k.
\end{aligned}
\end{equation}
\end{widetext}
The last term, $H_3(\boldsymbol{k})$, can be written in a similar way to $H_2(\bm{k})$. Therefore, the Hamiltonian in reciprocal space is
\begin{equation}
\begin{split}
H(\boldsymbol{k})=&-s'(\bm{k})a_{\bm{k}}^{\dagger}b_{\bm{k}}- \Delta s'(\bm{k}+p\bm{K}_+'+q\bm{K}_-')a_{\bm{k}+\bm{G}'}^{\dagger}b_{\bm{k}}\\
&-\Delta^* s'(\bm{k}-p\bm{K}_+'-q\bm{K}_-')a_{\bm{k}-\bm{G}'}^{\dagger}b_{\bm{k}}+H.c..
\end{split}
\end{equation}
This expression can be rewritten in terms of a $6\times6$ matrix, by defining the column vector $c_{\bm{k}}=(a_{\bm{k}}, a_{\bm{k}-\bm{G}'}, a_{\bm{k}+\bm{G}'}, b_{\bm{k}}, b_{\bm{k}-\bm{G}'}, b_{\bm{k}+\bm{G}'})$, resulting in,
\begin{equation}\label{Eq:H-k}
H(\bm{k}) = -c_{\bm{k}}^\dagger 
\begin{pmatrix}
0 & \Gamma \\
\Gamma^\dagger & 0 \\
\end{pmatrix}
c_{\bm{k}},
\end{equation}
where,
\begin{widetext}
\begin{equation}\label{Eq:Matrix-Gamma}
\Gamma =
\begin{pmatrix}
s'(\bm{k}) & \Delta s'(\bm k-\bm G'+p \bm K_+'+q \bm K_-') & \Delta^* s'(\bm k+ \bm G'-p \bm K_+'-q\bm K_-')\\
\Delta^* s'(\bm k-p \bm K_+'-q \bm K_-') & s'(\bm k- \bm G') & \Delta s'(\bm k+ \bm G'+p \bm K_+'+q \bm K_-') \\
\Delta s'(\bm k+p \bm K_+'+q \bm K_-') & \Delta^* s'(\bm k- \bm G'-p \bm K_+'-q \bm K_-') & s'(\bm k+ \bm G')\\
\end{pmatrix}
.
\end{equation}
\end{widetext}
Eq.~\cref{Eq:Matrix-Gamma} can be further simplified by using the relation 
\begin{equation}
s'(\bm{k}+p\bm{K}'_++q\bm{K}_-')=e^{i\frac{2 \pi}{3}(p+q)} s'(\bm{k}+(\nu-1)\bm{G}'),
\label{Eq:App.A}
\end{equation}
and defining $\tilde{\Delta}=e^{i\frac{2 \pi}{3}(p+q)} \Delta$ and $s'_n=s'(\bm k+n \bm G')$, to obtain, 
\begin{equation}
\Gamma = 
\begin{pmatrix}
s'_0 & \tilde{\Delta}s'_{\nu+1} & \tilde{\Delta}^* s'_{-\nu-1}\\
\tilde{\Delta}^* s'_{1-\nu} & s'_{-1} & \tilde{\Delta} s'_{\nu} \\
\tilde{\Delta} s'_{\nu-1} & \tilde{\Delta}^* s'_{-\nu} & s'_1\\
\end{pmatrix}
.\label{Eq:MGamma}
\end{equation}
As a result, the spectrum is symmetric around $E=0$ and is determined by $|\Gamma|^{2}=\Gamma^\dagger\Gamma$. To ilustrate this, we simply calculate
\begin{equation}
H^{2}(\bm{k})=-c_{\bm{k}}^\dagger 
\begin{pmatrix}
|\Gamma|^{2} & 0 \\
0 & |\Gamma|^{2} \\
\end{pmatrix}
c_{\bm{k}},
\end{equation}
with characteristic polynomial,
\begin{equation}
\det\left(|\Gamma|^{2}-E^{2}(\bm{k})\right)=0.
\end{equation}

\begin{figure}[!htbp]
\begin{center}
\includegraphics[width=4.25cm]{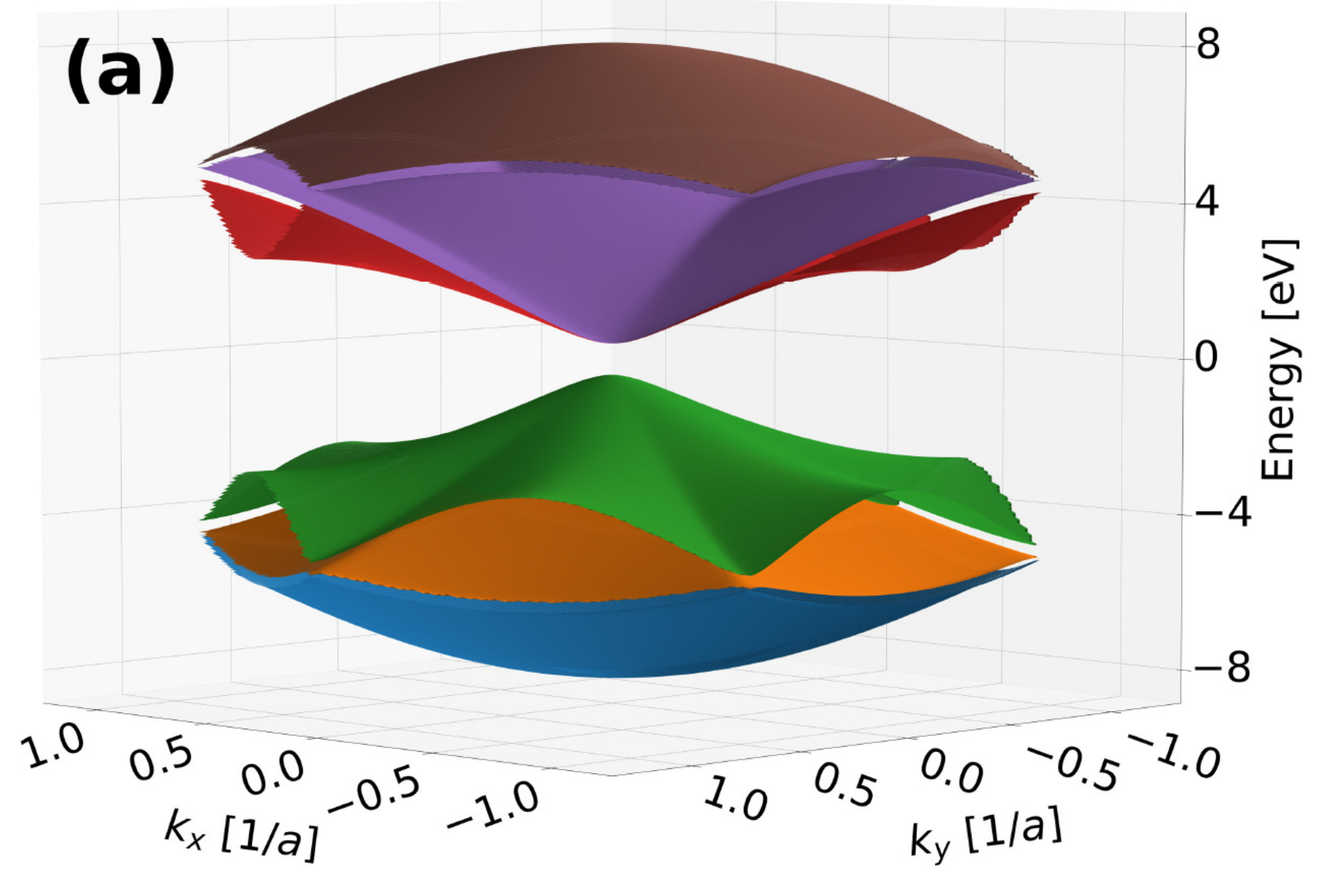}
\includegraphics[width=4.25cm]{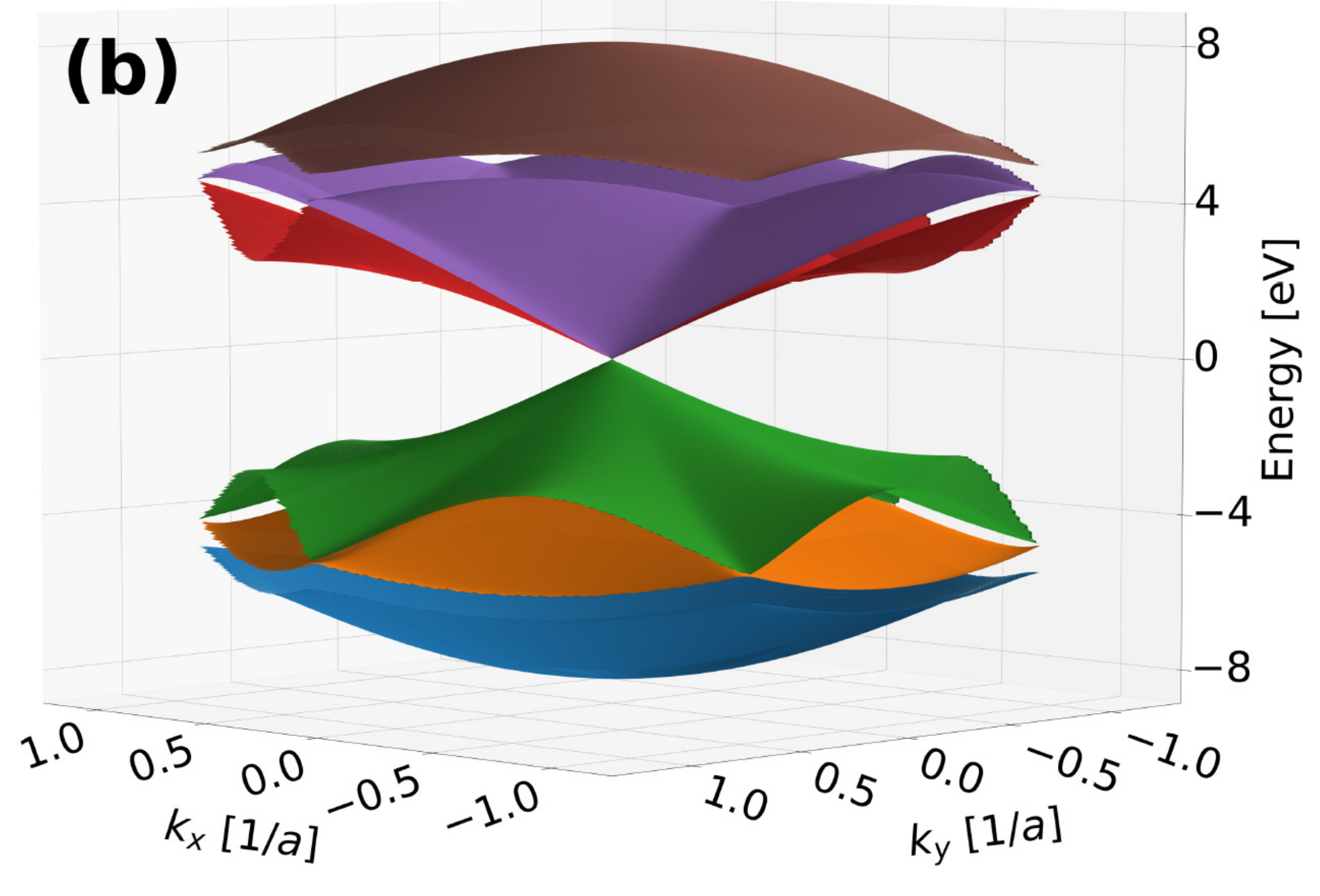}
\includegraphics[width=4.25cm]{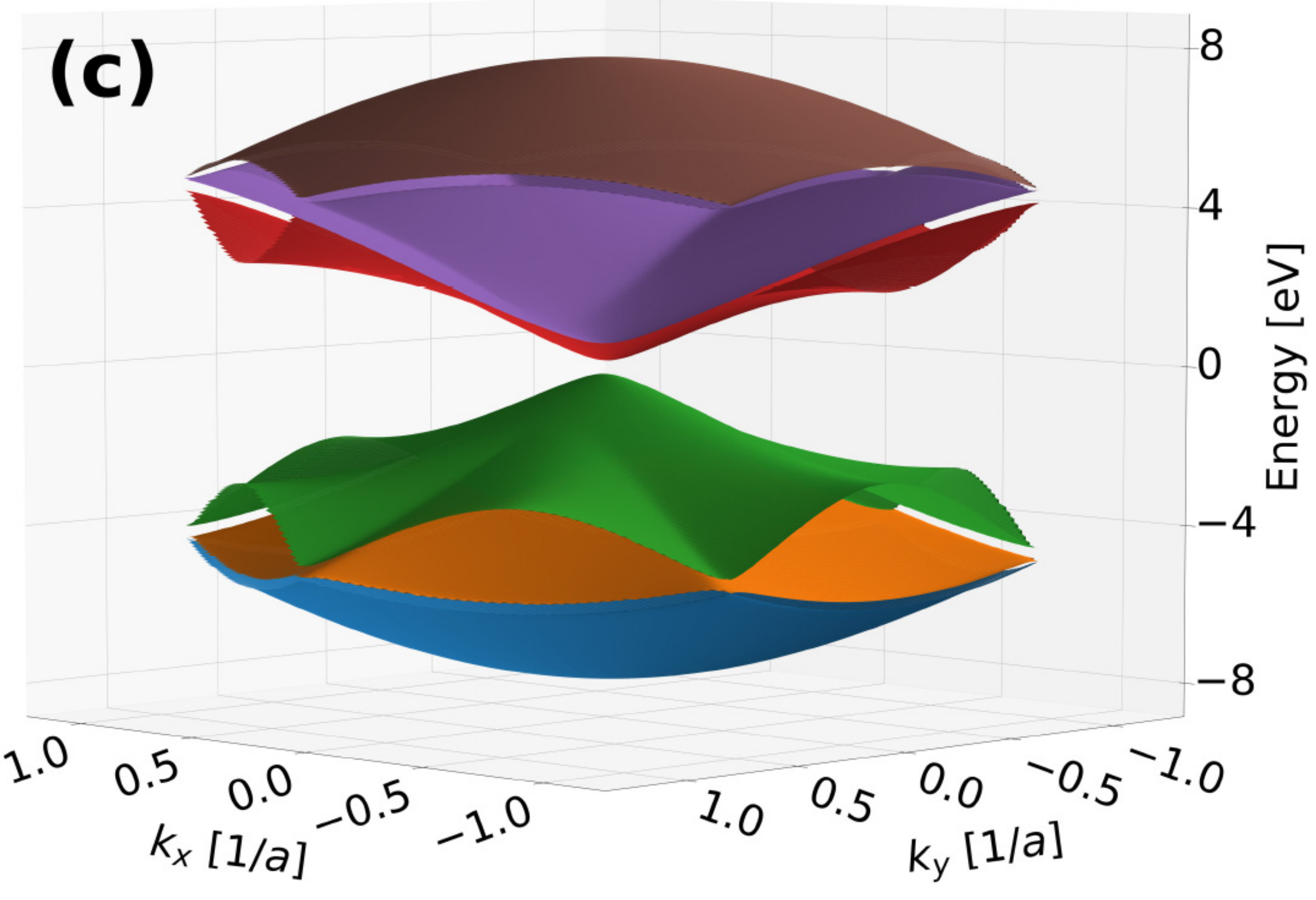}
\includegraphics[width=4.25cm]{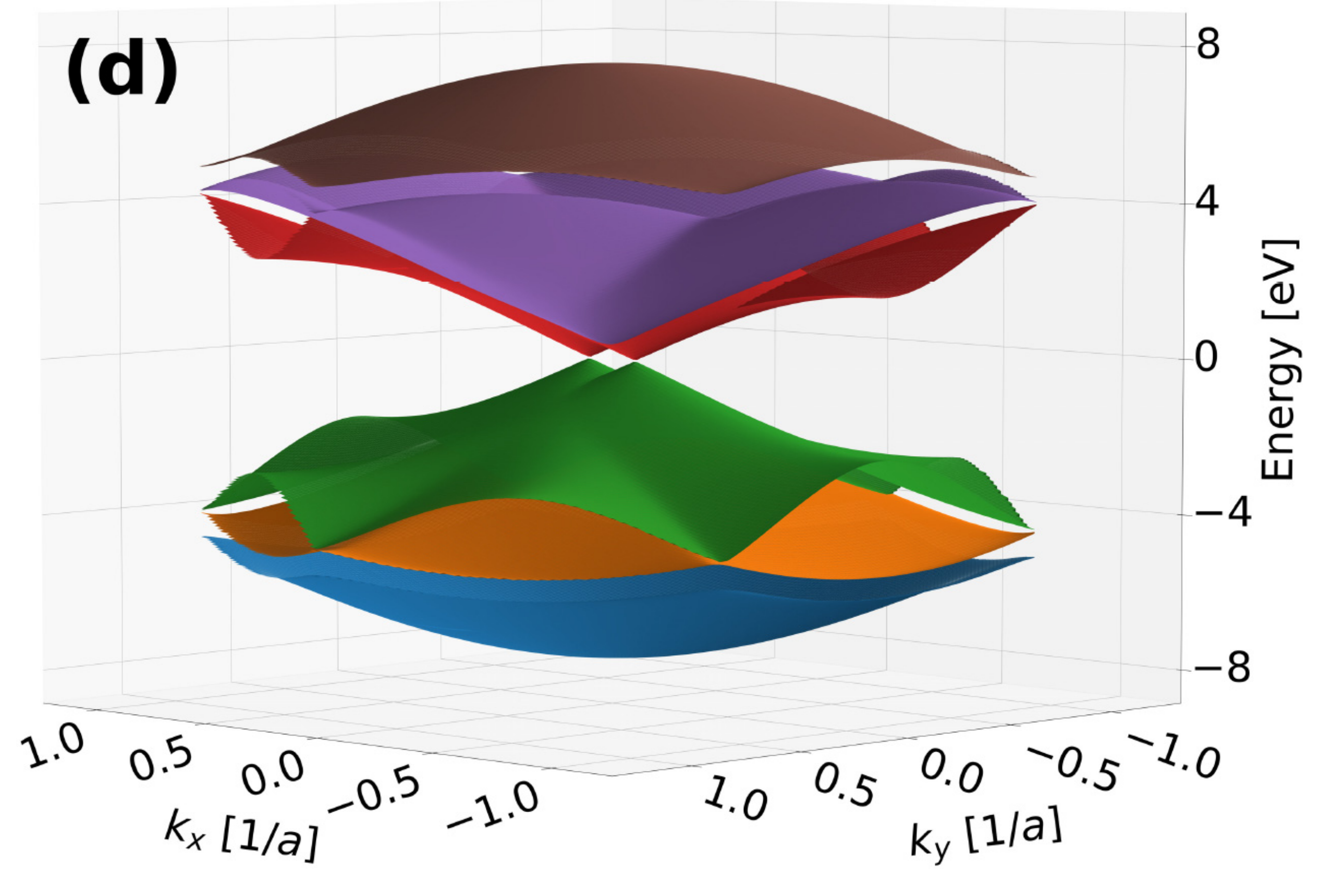}
\end{center}
\caption{Energy dispersion for, a) Kek-O graphene, b) Kek-Y Graphene, c) Kek-O graphene with uniform strain in the $x$ direction ($\theta=0$), d) Kek-Y graphene with uniform strain in the $x$ direction. The plots were produced using $\Delta=0.05$ for all figures, and $\epsilon=0.03$ for c) and $\epsilon=0.05$ for d).} \label{Fig:EnergyDispersion}
\end{figure}

In Figure \ref{Fig:EnergyDispersion} we show a comparison between the energy dispersions for (a) graphene with a Kek-O pattern, (b) graphene with a Kek-Y pattern, (c)  with a Kek-O pattern and strain, and (d)  with a Kek-Y pattern and strain.  From Fig.~\ref{Fig:EnergyDispersion}(a) and Fig.~\ref{Fig:EnergyDispersion}(c) it is clear that a gap is preserved for small values of $\epsilon$, although its size is considerable reduced when compared with the pure  Kek-O pattern. This results from a competition between the Kekul\'e parameter that opens a gap and the magnitude of strain which closes it. Once the gap is closed, an increase of the strain results in two shifted Dirac cones (not shown).

Figure \ref{Fig:EnergyDispersion}~(d) shows the results for a Kek-Y pattern with strain. Here, the effects of strain are much more important, as the central Dirac cones are no longer uniaxial, resulting in two separate Dirac cones.  For this phase, strain preserves the massless character and moves the cones away from the center of the Brillouin zone. In Figure \ref{fig:NuevaBrillouin} we provide a short pictorial summary of the Dirac cones' fate  after applying a pure Kekul\'e, strain, or a Kekul\'e plus strain modulations.  For the case of Kekul\'e plus strain, the tips of the two cones are much closer in reciprocal space than in the case of graphene.  This suggests that strain can be used to control the distance between valleys to do valley engineering. As the electrical conductivity\cite{Oliva2014} and the optical properties \cite{OlivaDicroism,Oliva2016_PRB}  depend upon the distance in k-space of the cone tips, it is clear that strain valley engineering can be much more effective in Kekul\'e patterns than in pure graphene. 
In the following section, we will consider a low energy approximation that allows us to obtain a useful effective Dirac Hamiltonian for this system. 

\section{Low-energy approximation}
In order to obtain an effective Hamiltonian for low energies,  we start by observing that the first row and column of the matrix $\Gamma$ given by Eq. \cref{Eq:MGamma} are negligible in such limit, since they correspond to the high energy bands depicted in brown and blue in Figure \ref{Fig:EnergyDispersion}. As a result, we can redefine the column vector of annihilation operators as $u_{\bm{k}}=(a_{\bm{k}-\bm{G}'}, a_{\bm{k}+\bm{G}'}, b_{\bm{k}-\bm{G}'}, b_{\bm{k}+\bm{G}'})$. The effective Hamiltonian now can be written as follows, 
\begin{equation}
H_{Eff}=-u_{\bm{k}}^{\dagger}
\begin{pmatrix}
0 & 0 & s'_{-1} & \tilde{\Delta} s'_{\nu} \\
0 & 0 & \tilde{\Delta}^* s'_{-\nu} & s'_1 \\
s'^*_{-1} & \tilde{\Delta} s'^*_{-\nu} & 0 & 0\\
\tilde{\Delta}^* s'^*_{\nu} & s'^*_1 & 0 & 0\\
\end{pmatrix}
u_{\bm{k}}.
\label{Eq:Matrix}
\end{equation}

\begin{figure}[!htbp]
\begin{center}
\includegraphics[width=8.5cm]{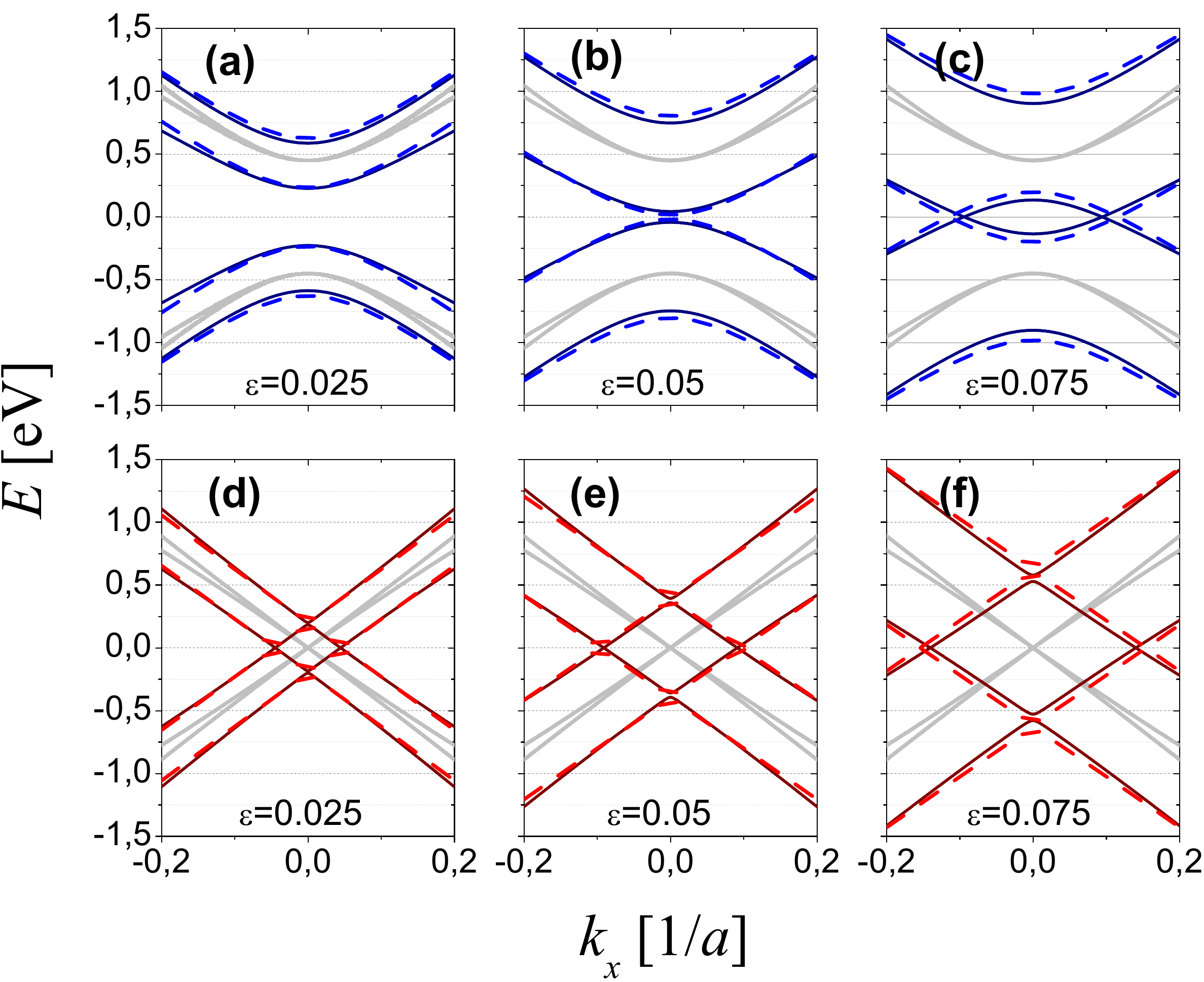} 
\end{center}
\caption{Energy dispersion relation around the $\Gamma$ point for Kekul\'e distorted graphene with parameter $\Delta=0.05$, and for different values of uniaxial strain along the $x$-axis ($\theta=0$). Panels (a-c) show the results for the Kek-O texture (blue curves) with $2.5\%$, $5\%$, and $7.5\%$ of strain, respectively. Panels (d-f) show (in red) the corresponding curves for the Kek-Y texture. The solid curves were obtained by solving numerically the Hamiltonian given by Eq.\cref{Eq:H-k}, while the dashed curves were calculated by using the analytical expressions for the low energy approximation given by Eq.~\cref{Eq:LowKek-O} and Eq.~\cref{Eq:LowKek-Y}. The gray lines corresponds to the unstrained cases. Notice the good agreement between the numerical and low-energy approximation}.
\label{fig:BandStructures}
\end{figure}

Next, we proceed to expand Eq.~\ref{Eq:Matrix} to first order in $\bm k$. To this end we can make an expansion of the energy dispersion $s'_n$ around $n \bm G'$. However, as other works have shown \cite{oliva-Leyva-PRB,NaumisReview}, it is necessary to expand around the true Dirac points, which are defined as the zeros of the deformed lattice energy dispersion, not located at the high-symmetry points of the strained-lattice, or at the original Dirac cones' tips. These new Dirac points are given by $ \bm K_D^{'\pm}=\pm (\bm G'+\bm A)$, where $\bm A$ is the pseudo-magnetic vector potential\cite{oliva-Leyva-PRB,NaumisReview}, whose explicit form depends upon the components of the strain tensor $\bar{\epsilon}$:
\begin{equation}
A_x=\frac{\beta}{2a}(\epsilon_{xx}-\epsilon_{yy}), \quad A_y=-\frac{\beta}{2a}(2 \epsilon_{xy}).
\end{equation}
By writing $s'_n$ as $s'(\bm k + n \bm G'+n \bm A-n \bm A)=s'([\bm k-n \bm A]+n\bm K_D^{'+})$ we can explicitly ensure that the expansion is performed around the true Dirac points. Then we return to $n \bm G'$ through a translation of $-n \bm A$, such that
\begin{equation}
s'_n \approx s'(n \bm K_D^{'+})+ \nabla_k s'(\bm k) \rvert_{\bm k= n \bm K_D^{'+}} \cdot (\bm k-n\bm A) + O(\bm k^2).\label{eq:expansion}
\end{equation}

Thus, the matrix elements of Eq. \cref{Eq:Matrix} can be expressed as follows:
\begin{subequations}
\begin{equation}
s'_0 \approx 3\tilde{t_0}+iv_f a (\bm A \times \bm p)_z,
\end{equation}
\begin{equation}
s'_1 \approx -v_f[(\overline{p}_x-\hbar A_x)-i[(\overline{p}_y-\hbar A_y)],
\end{equation}
\begin{equation}
s'_{-1} \approx v_f[(\overline{p}_x+\hbar A_x)+i(\overline{p}_y+\hbar A_y)],
\end{equation}
\end{subequations}
where $\tilde{t_0}=t_0[1-\frac{\beta \epsilon}{2}(1-\rho)]$, $\overline{\bm p}=(1+\bar{\epsilon}-\beta \bar{\epsilon}) \bm p$, 
and we defined the Fermi velocity $v_f=3at_0/2 \hbar$ as usual.
Finally, we can write the Dirac-like equation for electrons in strained graphene with a Kekul\'e distorsion as, 
\begin{subequations}
\begin{equation}
\mathcal{H}
\begin{pmatrix}
\Psi_- \\
\Psi_+
\end{pmatrix}
=E
\begin{pmatrix}
\Psi_- \\
\Psi_+
\end{pmatrix}
,
\end{equation}
\begin{equation}
\Psi_-=
\begin{pmatrix}
-\psi_{B,-} \\
\psi_{A,-} \\
\end{pmatrix}
, \quad
\Psi_{+}=
\begin{pmatrix}
\psi_{A,+} \\
\psi_{B,+} \\
\end{pmatrix}
,
\end{equation}

\begin{equation}\label{Eq:H-cal}
\mathcal{H}=
\begin{pmatrix}
v_f \bm \sigma \cdot (\overline{\bm p}+ \hbar \bm A) & \tilde{\Delta}Q_{\nu}\\
 \tilde{\Delta}^*Q_{\nu}^ \dagger & v_f \bm \sigma \cdot (\overline{\bm p}- \hbar \bm A)
\end{pmatrix}
,
\end{equation}

\begin{equation} 
Q_{\nu}= 
\left \{
\begin{matrix}
3\tilde{t_0}\sigma_z-iv_fa(\bm A \times \bm p)_z \sigma_0, & \nu=0\\
v_f[\nu \overline{p}_x- i \overline{p}_y]\sigma_0 +\hbar v_f[A_x- i \nu A_y]\sigma_z, &\nu=\pm 1
\end{matrix}
\right.,
\end{equation}
\end{subequations}
where $\psi_{s,v}$ is the wavefunction for sublattice $s\in\{A,B\}$ in the valley  $v\in\{+,-\}$ and the Pauli matrices $\sigma_i$, $i\in\{0,x,y,z\}$ are acting in the pseudospin degree of freedom.

The energy eigenvalues of the Hamiltonian \cref{Eq:H-cal} can be obtained analytically for both Kekul\'e textures.
For the Kek-O texture ($\nu=0$), we obtain the following expression,
\begin{widetext}
\begin{equation}
\begin{aligned}
E_{O,\pm}^2=&v_f^2(|\overline{\bm p}|^2 +\hbar^2 \bm |A|^2) +\Delta_0^2([ 3\tilde{t_0}]^2+[v_fa(\bm A \times \bm p)_z]^2)\\
&\pm 2 v_f \: \sqrt[]{v_f^2 \hbar^2 (\bm A \cdot \overline{\bm p})^2+\Delta_0^2 (v_f^2 a^2 |\overline{\bm p}|^2(\bm A \times \bm p)_z^2+2v_f\hbar a(3\tilde{t_0})(\bm A \times \bm p)_z(\bm A \times \overline{\bm p})_z+\hbar^2 (\tilde{3t_0})^2 |\bm A|^2)},
\label{Eq:LowKek-O}
\end{aligned}
\end{equation}
\end{widetext}
which recovers the result for the Kek-O unstrained graphene\cite{Gamayun} and may be evaluated at ($k_x,k_y$)=($0,0$) to find the condition for keeping the gap\cite{Park2015},
\begin{equation}\label{Eq:gapconditon}
\epsilon < \frac{4 |\Delta_0|}{\beta[(1+\rho)+2 |\Delta_0|(1-\rho)]} 
\end{equation}
as well as its magnitude,
\begin{equation}
E_{Gap}=6 t_0[|\Delta_0|-\frac{\beta \epsilon}{4}(1+\rho)-\frac{\beta \epsilon}{2}(1-\rho)|\Delta_0|]
\end{equation}
This characterizes the competition  between the Kekul\'e strength $\Delta_0$, and the magnitude of the applied strain $\epsilon$ to open and close the gap, and suggest a way to control this gap by strain. It has been pointed out that this gap becomes topological\cite{wakabayashi} for negative values of $\Delta_0$. Notice that both equations are independent from the strain direction $\theta$, this is a consequence of the approximations made for small strain and low-energy. When the magnitude of strain equals the condition given by Eq.~\cref{Eq:gapconditon}, the gap closes and the valence and conduction bands touches in just one point. For greater values of strain, the bands touch in two points (valleys) that split as strain increases. This is shown in the series of plots in Fig.~\ref{fig:BandStructures}(a)-\ref{fig:BandStructures}(b), where the dispersion relations around the center of the Brillouin zone for strained Kek-O graphene are presented. Gray lines are the dispersions for unstrained graphene with Kek-O texture. Blue continuous lines present the curves obtained numerically by calculating  the eigenvalues of Eq.\cref{Eq:H-k} while dashed lines are the energies in the low energy approximation of Eq.\cref{Eq:LowKek-O}. They correspond to the eigenvalues of the effective Hamiltonian for the Kek-O texture given by,
\begin{equation}
\begin{aligned}
\mathcal{H}_O=&v_f \bm \sigma \cdot \overline{\bm p}  \otimes \tau_0+\hbar v_f \bm \sigma \cdot \bm A \otimes \tau_z\\
&+ 3\Delta_0 t_0 \sigma_z \otimes \tau_x    + v_{\tau}a (\bm A \times \bm p)_z \sigma_0 \otimes \tau_y,
\end{aligned}
\end{equation}
where we have taken $\tilde{\Delta}=\Delta_0$, and used a second set of Pauli matrices $\tau_x$, $\tau_y$, $\tau_z$, with a unit matrix $\tau_0$ acting on valley space and defined the velocity $v_{\tau}=\Delta_0 v_{f}$. Notice that the pseudomagnetic vector $\bm{A}$, as usual, appears as a momentum shift, nevertheless since it does not depend on space it can not give rise to a pseudomagnetic field\cite{vozmediano2010gauge,AMORIM20161,NaumisReview}.

For the Kek-Y texture ($\nu=\pm 1$), a gapless spectrum remains for all values of Keukul\'e and strain parameters, with energies given by,
\begin{widetext}
\begin{equation}
\begin{aligned}
E_{Y,\pm}^2=&v_f^2(1+\Delta_0^2)(|\overline {\bm p}|^2+\hbar^2 |\bm A|^2)\\
& \pm \frac{v_f^2}{2} \, \sqrt[]{(1+\Delta_0^2)^2(|\overline{\bm p} + \hbar \bm A|^4+|\overline{\bm p} - \hbar \bm A|^4)-2|\overline{\bm p} + \hbar \bm A|^2|\overline{\bm p} - \hbar \bm A|^2(1-6\Delta_0^2+\Delta_0^4)}. 
\label{Eq:LowKek-Y}
\end{aligned}
\end{equation}
\end{widetext}

In Fig. \ref{fig:BandStructures} we present a comparison between  Eq.~\cref{Eq:LowKek-O} and Eq.~\cref{Eq:LowKek-Y}, with a calculation obtained from a numerical diagonalization of Eq. \cref{Eq:H-k}. The good agreement between both calculations validate our expressions for low-energy.

Finally, by taking $\nu=1$ and $\tilde{\Delta}=\Delta_0$, we obtain the low-energy effective Hamiltonian for the Kek-Y texture, 
\begin{equation}
\begin{aligned}
\mathcal{H}_Y=&v_f \bm \sigma \cdot \overline{\bm p}  \otimes \tau_0+\hbar v_f \bm \sigma \cdot \bm A \otimes \tau_z\\
&+v_{\tau} \sigma_0 \otimes \bm \tau \cdot \overline{\bm p}    + \hbar v_{\tau} \sigma_z \otimes \bm \tau \cdot \bm A.
\end{aligned}
\end{equation}
An equivalent expression is found for $\nu=-1$ and a complex $\tilde{\Delta}$.

When compared with the pure Kekul\'e effective Hamiltonian \cite{Gamayun}, we observe two new terms, both containing $\bm{A}$. These two terms are $k$-independent and have a Zeeman-like structure, one in the pseudospin quantum number as it contains the product $\bm{\sigma} \cdot \bm{A}$, and the other in the valley quantum number  (proportional to $\bm{\tau} \cdot \bm{A}$). The former is the leading term, since it depends linearly on $\epsilon$, while the latter depends on the product of $\epsilon\Delta$. Since the first term contains $\tau_z$, it splits the two valleys by moving each cone in opposite directions away from the center $\Gamma$ of the superlattice Brillouin zone, as shown in Fig.\ref{fig:BandStructures}~(d-f). The second term has a similar effect but in pseudospin space, nevertheless for modest values of strain and Kekul\'e distorsion it can be neglected. 
Although the first term is proportional to $\tau_z$ and the second is proportional to $\sigma_z$, both preserve the valley and pseudospin energy degeneracy.

\section{Conclusions}

We studied the effects upon the electronic properties of a space-independent strain in different types of Kekul\'e-patterns in graphene. For  the Kek-O type, moderated values of strain preserve the gap although the size is changed. Above a certain strain threshold value, the gap closes leaving a two-Dirac-cones dispersion. For  the Kek-Y type, strain splits the valleys along the direction of applied strain. However, as the valleys were folded before by the Kekul\'e pattern, it turns out that the distance in reciprocal space of the valleys is much closer than in pure graphene. This suggest that strain is useful to control the degree of intervalley scattering in Kekul\'e patterns. We also provided a low-energy Dirac effective Hamiltonian, which presents a Zeeman-like coupling between pseudospin and valleys to the pseudomagnetic vectorial potential.

\section*{Acknowledgments}
E.A. and R.C.-B. acknowledges useful discussions with Francisco Mireles, Pierre A. Pantaleon, Mahmoud Asmar and David Ruiz Tijerina. The plots in Fig.\ref{fig:BandStructures} were created using the software Kwant \citep{groth2014kwant}. This work was supported by project UNAM-DGAPA-PAPIIT-IN102717.

\bibliographystyle{unsrt}
\bibliography{refs.bib}

\begin{thebibliography}{10}

\bibitem{Schaibley}
John~R Schaibley, Hongyi Yu, Genevieve Clark, Pasqual Rivera, Jason~S Ross,
  Kyle~L Seyler, Wang Yao, and Xiaodong Xu.
\newblock Valleytronics in 2d materials.
\newblock {\em Nature Reviews Materials}, 1(11):16055, 2016.

\bibitem{neto2009electronic}
AH~Castro Neto, F~Guinea, Nuno~MR Peres, Kostya~S Novoselov, and Andre~K Geim.
\newblock The electronic properties of graphene.
\newblock {\em Reviews of modern physics}, 81(1):109, 2009.

\bibitem{Chamon2007}
Chang-Yu Hou, Claudio Chamon, and Christopher Mudry.
\newblock Electron fractionalization in two-dimensional graphenelike
  structures.
\newblock {\em Phys. Rev. Lett.}, 98:186809, May 2007.

\bibitem{Chamon2000}
Claudio Chamon.
\newblock Solitons in carbon nanotubes.
\newblock {\em Phys. Rev. B}, 62:2806--2812, Jul 2000.

\bibitem{wakabayashi}
Feng Liu, Minori Yamamoto, and Katsunori Wakabayashi.
\newblock Topological edge states of honeycomb lattices with zero berry
  curvature.
\newblock {\em Journal of the Physical Society of Japan}, 86(12):123707, 2017.

\bibitem{wu2016}
Long-Hua Wu and Xiao Hu.
\newblock Topological properties of electrons in honeycomb lattice with detuned
  hopping energy.
\newblock {\em Scientific reports}, 6:24347, 2016.

\bibitem{manoharan}
Kenjiro~K Gomes, Warren Mar, Wonhee Ko, Francisco Guinea, and Hari~C Manoharan.
\newblock Designer dirac fermions and topological phases in molecular graphene.
\newblock {\em Nature}, 483(7389):306, 2012.

\bibitem{li-phonons}
Yizhou Liu, Chao-Sheng Lian, Yang Li, Yong Xu, and Wenhui Duan.
\newblock Pseudospins and topological effects of phonons in a kekul\'e lattice.
\newblock {\em Phys. Rev. Lett.}, 119:255901, Dec 2017.

\bibitem{ontop}
Zhuonan Lin, Wei Qin, Jiang Zeng, Wei Chen, Ping Cui, Jun-Hyung Cho, Zhenhua
  Qiao, and Zhenyu Zhang.
\newblock Competing gap opening mechanisms of monolayer graphene and graphene
  nanoribbons on strong topological insulators.
\newblock {\em Nano Letters}, 17(7):4013--4018, 2017.
\newblock PMID: 28534404.

\bibitem{Sorella}
Sandro Sorella, Kazuhiro Seki, Oleg~O. Brovko, Tomonori Shirakawa, Shohei
  Miyakoshi, Seiji Yunoki, and Erio Tosatti.
\newblock Correlation-driven dimerization and topological gap opening in
  isotropically strained graphene.
\newblock {\em Phys. Rev. Lett.}, 121:066402, Aug 2018.

\bibitem{cheianov}
V.V. Cheianov, V.I. Fal’ko, O.~Syljuåsen, and B.L. Altshuler.
\newblock Hidden kekulé ordering of adatoms on graphene.
\newblock {\em Solid State Communications}, 149(37):1499 -- 1501, 2009.

\bibitem{Gutierrez}
Christopher Guti{\'e}rrez, Cheol-Joo Kim, Lola Brown, Theanne Schiros, Dennis
  Nordlund, Edward~B Lochocki, Kyle~M Shen, Jiwoong Park, and Abhay~N
  Pasupathy.
\newblock Imaging chiral symmetry breaking from kekul{\'e} bond order in
  graphene.
\newblock {\em Nature Physics}, 12(10):950, 2016.

\bibitem{Gamayun}
O~V Gamayun, V~P Ostroukh, N~V Gnezdilov, İ~Adagideli, and C~W~J Beenakker.
\newblock Valley-momentum locking in a graphene superlattice with y-shaped
  kekulé bond texture.
\newblock {\em New Journal of Physics}, 20(2):023016, 2018.

\bibitem{rycerz2007valley}
A~Rycerz, J~Tworzyd{\l}o, and CWJ Beenakker.
\newblock Valley filter and valley valve in graphene.
\newblock {\em Nature Physics}, 3(3):172, 2007.

\bibitem{Fujita2010}
T.~Fujita, M.~B.~A. Jalil, and S.~G. Tan.
\newblock Valley filter in strain engineered graphene.
\newblock {\em Applied Physics Letters}, 97(4):043508, 2010.

\bibitem{guinea2013}
Yongjin Jiang, Tony Low, Kai Chang, Mikhail~I. Katsnelson, and Francisco
  Guinea.
\newblock Generation of pure bulk valley current in graphene.
\newblock {\em Phys. Rev. Lett.}, 110:046601, Jan 2013.

\bibitem{wang2014}
J.~Wang and S.~Fischer.
\newblock Topological valley resonance effect in graphene.
\newblock {\em Phys. Rev. B}, 89:245421, Jun 2014.

\bibitem{grujic2014}
Marko~M. Gruji\ifmmode~\acute{c}\else \'{c}\fi{}, Milan \ifmmode
  \check{Z}\else~\v{Z}\fi{}. Tadi\ifmmode~\acute{c}\else \'{c}\fi{}, and
  Fran\ifmmode \mbox{\c{c}}\else \c{c}\fi{}ois~M. Peeters.
\newblock Spin-valley filtering in strained graphene structures with
  artificially induced carrier mass and spin-orbit coupling.
\newblock {\em Phys. Rev. Lett.}, 113:046601, Jul 2014.

\bibitem{ren2015}
Yafei Ren, Xinzhou Deng, Zhenhua Qiao, Changsheng Li, Jeil Jung, Changgan Zeng,
  Zhenyu Zhang, and Qian Niu.
\newblock Single-valley engineering in graphene superlattices.
\newblock {\em Phys. Rev. B}, 91:245415, Jun 2015.

\bibitem{carrillo2016strained}
R~Carrillo-Bastos, C~Le{\'o}n, D~Faria, A~Latg{\'e}, Eva~Y Andrei, and
  N~Sandler.
\newblock Strained fold-assisted transport in graphene systems.
\newblock {\em Physical Review B}, 94(12):125422, 2016.

\bibitem{stegmann2016valley}
Thomas Stegmann and Nikodem Szpak.
\newblock Current flow paths in deformed graphene: from quantum transport to
  classical trajectories in curved space.
\newblock {\em New Journal of Physics}, 18(5):053016, 2016.

\bibitem{jones2017quantized}
Gareth~W Jones, Dario~Andres Bahamon, Antonio~H Castro~Neto, and Vitor~M
  Pereira.
\newblock Quantized transport, strain-induced perfectly conducting modes, and
  valley filtering on shape-optimized graphene corbino devices.
\newblock {\em Nano letters}, 17(9):5304--5313, 2017.

\bibitem{wu2016full}
Qing-Ping Wu, Zheng-Fang Liu, Ai-Xi Chen, Xian-Bo Xiao, and Zhi-Min Liu.
\newblock Full valley and spin polarizations in strained graphene with rashba
  spin orbit coupling and magnetic barrier.
\newblock {\em Scientific reports}, 6, 2016.

\bibitem{Cazalilla}
Xian-Peng Zhang, Chunli Huang, and Miguel~A Cazalilla.
\newblock Valley hall effect and nonlocal transport in strained graphene.
\newblock {\em 2D Materials}, 4(2):024007, 2017.

\bibitem{Asmar-minimal}
Mahmoud~M. Asmar and Sergio~E. Ulloa.
\newblock Minimal geometry for valley filtering in graphene.
\newblock {\em Phys. Rev. B}, 96:201407, Nov 2017.

\bibitem{Luo2017}
Kun Luo, Tao Zhou, and Wei Chen.
\newblock Probing the valley filtering effect by andreev reflection in a zigzag
  graphene nanoribbon with a ballistic point contact.
\newblock {\em Phys. Rev. B}, 96:245414, Dec 2017.

\bibitem{Beenakker}
C.~W.~J. Beenakker, N.~V. Gnezdilov, E.~Dresselhaus, V.~P. Ostroukh,
  Y.~Herasymenko, \ifmmode \dot{I}\else~\.{I}\fi{}. Adagideli, and
  J.~Tworzyd\l{}o.
\newblock Valley switch in a graphene superlattice due to pseudo-andreev
  reflection.
\newblock {\em Phys. Rev. B}, 97:241403, Jun 2018.

\bibitem{Yee2017}
Yee~Sin Ang, Shengyuan~A. Yang, C.~Zhang, Zhongshui Ma, and L.~K. Ang.
\newblock Valleytronics in merging dirac cones: All-electric-controlled valley
  filter, valve, and universal reversible logic gate.
\newblock {\em Phys. Rev. B}, 96:245410, Dec 2017.

\bibitem{Brown2018}
Rory Brown, Niels~R. Walet, and Francisco Guinea.
\newblock Edge modes and nonlocal conductance in graphene superlattices.
\newblock {\em Phys. Rev. Lett.}, 120:026802, Jan 2018.

\bibitem{settnes2017valley}
Mikkel Settnes, Jos{\'e}~Hugo Garc{\'\i}a, and Stephan Roche.
\newblock Valley-polarized quantum transport generated by gauge fields in
  graphene.
\newblock {\em arXiv preprint arXiv:1705.09085}, 2017.

\bibitem{milovanovic2016strained}
SP~Milovanovi{\'c} and FM~Peeters.
\newblock Strained graphene hall bar.
\newblock {\em Journal of Physics: Condensed Matter}, 29(7):075601, 2016.

\bibitem{Roche}
Mikkel Settnes, Jose~H Garcia, and Stephan Roche.
\newblock Valley-polarized quantum transport generated by gauge fields in
  graphene.
\newblock {\em 2D Materials}, 4(3):031006, 2017.

\bibitem{carrillo2018enhanced}
Ramon Carrillo-Bastos, Marysol Ochoa, Sa{\'u}l~A Zavala, and Francisco Mireles.
\newblock Enhanced asymmetric valley scattering by scalar fields in non-uniform
  out-of-plane deformations in graphene.
\newblock {\em arXiv preprint arXiv:1806.04708}, 2018.

\bibitem{stegmann2018}
Thomas Stegmann and Nikodem Szpak.
\newblock Current splitting and valley polarization in elastically deformed
  graphene.
\newblock {\em arXiv preprint arXiv:1806.09576}, 2018.

\bibitem{zhai2018local}
Dawei Zhai and Nancy Sandler.
\newblock Local versus extended deformed graphene geometries for valley
  filtering.
\newblock {\em arXiv preprint arXiv:1806.11251}, 2018.

\bibitem{ValleyPRL2018}
Shu-guang Cheng, Haiwen Liu, Hua Jiang, Qing-Feng Sun, and X.~C. Xie.
\newblock Manipulation and characterization of the valley-polarized topological
  kink states in graphene-based interferometers.
\newblock {\em Phys. Rev. Lett.}, 121:156801, Oct 2018.

\bibitem{Xiao2007}
Di~Xiao, Wang Yao, and Qian Niu.
\newblock Valley-contrasting physics in graphene: Magnetic moment and
  topological transport.
\newblock {\em Phys. Rev. Lett.}, 99:236809, Dec 2007.

\bibitem{xiao-review}
Di~Xiao, Ming-Che Chang, and Qian Niu.
\newblock Berry phase effects on electronic properties.
\newblock {\em Rev. Mod. Phys.}, 82:1959--2007, Jul 2010.

\bibitem{AMORIM20161}
B.~Amorim, A.~Cortijo, F.~de~Juan, A.G. Grushin, F.~Guinea,
  A.~Gutiérrez-Rubio, H.~Ochoa, V.~Parente, R.~Roldán, P.~San-Jose,
  J.~Schiefele, M.~Sturla, and M.A.H. Vozmediano.
\newblock Novel effects of strains in graphene and other two dimensional
  materials.
\newblock {\em Physics Reports}, 617(Supplement C):1 -- 54, 2016.
\newblock Novel effects of strains in graphene and other two dimensional
  materials.

\bibitem{NaumisReview}
Gerardo~G Naumis, Salvador Barraza-Lopez, Maurice Oliva-Leyva, and Humberto
  Terrones.
\newblock Electronic and optical properties of strained graphene and other
  strained 2d materials: a review.
\newblock {\em Reports on Progress in Physics}, 80(9):096501, 2017.

\bibitem{vozmediano2010gauge}
Mar{\'\i}a~AH Vozmediano, MI~Katsnelson, and Francisco Guinea.
\newblock Gauge fields in graphene.
\newblock {\em Physics Reports}, 496(4):109--148, 2010.

\bibitem{gorbachev2014}
RV~Gorbachev, JCW Song, GL~Yu, AV~Kretinin, F~Withers, Y~Cao, A~Mishchenko,
  IV~Grigorieva, KS~Novoselov, LS~Levitov, et~al.
\newblock Detecting topological currents in graphene superlattices.
\newblock {\em Science}, 346(6208):448--451, 2014.

\bibitem{komatsu2018}
Katsuyosih Komatsu, Yoshifumi Morita, Eiichiro Watanabe, Daiju Tsuya, Kenji
  Watanabe, Takashi Taniguchi, and Satoshi Moriyama.
\newblock Observation of the quantum valley hall state in ballistic graphene
  superlattices.
\newblock {\em Science Advances}, 4(5):eaaq0194, 2018.

\bibitem{Crommie}
N.~Levy, S.~A. Burke, K.~L. Meaker, M.~Panlasigui, A.~Zettl, F.~Guinea,
  A.~H.~Castro Neto, and M.~F. Crommie.
\newblock Strain-induced pseudo{\textendash}magnetic fields greater than 300
  tesla in graphene nanobubbles.
\newblock {\em Science}, 329(5991):544--547, 2010.

\bibitem{Theory-LL01}
Francisco Guinea, MI~Katsnelson, and AK~Geim.
\newblock Energy gaps and a zero-field quantum hall effect in graphene by
  strain engineering.
\newblock {\em Nature physics}, 6(1):30--33, 2010.

\bibitem{Theory-LL02}
F.~Guinea, A.~K. Geim, M.~I. Katsnelson, and K.~S. Novoselov.
\newblock Generating quantizing pseudomagnetic fields by bending graphene
  ribbons.
\newblock {\em Phys. Rev. B}, 81:035408, Jan 2010.

\bibitem{guinea}
L.~Gonz\'alez-\'Arraga, F.~Guinea, and P.~San-Jose.
\newblock Modulation of kekul\'e adatom ordering due to strain in graphene.
\newblock {\em Phys. Rev. B}, 97:165430, Apr 2018.

\bibitem{oliva-Leyva-PRB}
M.~Oliva-Leyva and Gerardo~G. Naumis.
\newblock Understanding electron behavior in strained graphene as a reciprocal
  space distortion.
\newblock {\em Phys. Rev. B}, 88:085430, Aug 2013.

\bibitem{Pereira2009}
Vitor~M. Pereira, A.~H. Castro~Neto, and N.~M.~R. Peres.
\newblock Tight-binding approach to uniaxial strain in graphene.
\newblock {\em Phys. Rev. B}, 80:045401, Jul 2009.

\bibitem{oliva-leyva-physA}
M.~Oliva-Leyva and Gerardo~G. Naumis.
\newblock Generalizing the fermi velocity of strained graphene from uniform to
  nonuniform strain.
\newblock {\em Physics Letters A}, 379(40):2645 -- 2651, 2015.

\bibitem{Botello2018}
Andr\'{e}s~R. Botello-M\'{e}ndez, Juan~Carlos Obeso-Jureidini, and Gerardo~G.
  Naumis.
\newblock Toward an accurate tight-binding model of graphene’s electronic
  properties under strain.
\newblock {\em The Journal of Physical Chemistry C}, 122(27):15753--15760,
  2018.

\bibitem{Oliva2014}
M~Oliva-Leyva and Gerardo~G Naumis.
\newblock Anisotropic ac conductivity of strained graphene.
\newblock {\em Journal of Physics: Condensed Matter}, 26(12):125302, 2014.

\bibitem{OlivaDicroism}
M~Oliva-Leyva and Gerardo~G Naumis.
\newblock Tunable dichroism and optical absorption of graphene by strain
  engineering.
\newblock {\em 2D Materials}, 2(2):025001, 2015.

\bibitem{Oliva2016_PRB}
M.~Oliva-Leyva and Gerardo~G. Naumis.
\newblock Effective dirac hamiltonian for anisotropic honeycomb lattices:
  Optical properties.
\newblock {\em Phys. Rev. B}, 93:035439, Jan 2016.

\bibitem{Park2015}
Joon-Suh Park and Hyoung~Joon Choi.
\newblock Band-gap opening in graphene: A reverse-engineering approach.
\newblock {\em Phys. Rev. B}, 92:045402, Jul 2015.

\bibitem{groth2014kwant}
Christoph~W Groth, Michael Wimmer, Anton~R Akhmerov, and Xavier Waintal.
\newblock Kwant: a software package for quantum transport.
\newblock {\em New Journal of Physics}, 16(6):063065, 2014.

\end{thebibliography}
\end{document}